\documentclass{intech}

\usepackage{ifsym}
\usepackage{bbding}
\usepackage{amsmath}

\chaptertitle{Electronic transport properties of few-layer graphene materials} 

\authors{S. Russo$^1$, M. F. Craciun$^1$, T. Khodkov$^1$}
\affiliation{$^1$ Centre for Graphene Science, College of Engineering, Mathematics and Physical Sciences, University of Exeter, Exeter} \country{United Kingdom}

\secondauthors{M. Koshino$^2$, M Yamamoto$^3$ and S Tarucha$^{3}$}
\secondaffiliation{$^2$Department of Physics, Tohoku University, Sendai

$^3$Department of Applied Physics, The
University of Tokyo, Tokyo}
\secondcountry{Japan}

\begin{document}

\maketitle

\section{Introduction} 

Since the discovery of graphene -a single layer of carbon atoms arranged in a honeycomb lattice - it was clear that this truly is a unique material system with an unprecedented combination of physical properties (\cite{Novoselov04,Novoselov05,Novoselov005,Zhang05,Geim07}). Graphene is the thinnest membrane present in nature -just one atom thick- it is the strongest material (\cite{CastroNeto09}), it is transparent (\cite{Nair08}) and it is a very good conductor (\cite{Novoselov05,Novoselov005}) with room temperature charge mobilities larger than the typical mobilities found in silicon. The significance played by this new material system is even more apparent when considering that graphene is the thinnest member of a larger family: the few-layer graphene materials (FLGs). Even though several physical properties are shared between graphene and its few-layers, recent theoretical and experimental advances demonstrate that each specific thickness of few-layer graphene is a material with unique physical properties (\cite{Nair08,Kim09,Bae10,Lee08}).

All few layers graphene are good conductors. However, striking differences in the nature of this conductive state emerge when a perpendicular electric field generated by gate voltages is applied onto the few-layers. In a single layer graphene transistor, the current is modulated by a gate voltage but it cannot be switched off since in the energy dispersion of graphene there is no band-gap (valence and conduction bands touch each other) (\cite{Novoselov05,Novoselov005}). Recent experimental advances showed that bilayer graphene, also characterized by touching valence and conduction bands,  develops an energy gap when subjected to an external perpendicular electric field (\cite{Zhang09}). Bilayer graphene is the only known semiconductor with a gate tuneable band-gap. Opposed to the case of single- and bi-layer, the trilayer material is a semimetal with a gate-tuneable band overlap between the conduction and the valence band. Indeed, the conductivity of trilayers increases when a perpendicular electric field is applied onto the system (\cite{Craciun09}). The variety of physical properties found in different FLGs is the true strength of these newly discovered materials, and yet very little is known on few-layer graphene with more than 3 layers.

\section{Electronic properties of few-layer graphene materials}\label{Electronic properties of FLGs}


The electronic properties of a material are intimately related to
its energy dispersion. There are several
approaches to calculate the electronic energy bands and
here we review the current understanding of the graphene materials band structure within the non-interacting tight-binding approximation (\cite{Wallace47}).
When two or more carbon atoms are brought together to form a
regular lattice -such as the hexagonal lattice for a single layer
graphene- the valence electrons of the different atoms interact.
This leads to a broadening of the electronic eigenstates and
ultimately to the formation of the continuous bands of a solid. 

An isolated carbon atom has 6 electrons $1s^2
2s^2 2p^2$, where the energies of the s-orbital and p-orbitals of
the second electronic shell are very similar. Consequently, carbon
can form a number of hybridized atomic orbitals characterizing
different geometries. In the case of graphene, one s-orbital and
two p-orbitals ($p_x$ and $p_y$) undergo a sp2 hybridization with
a characteristic planar trigonal symmetry with an angle of 120\textdegree between each bond. This is the reason why each carbon atom within
graphene has three nearest neighbors at a distance of $a_{0}=0.142 nm$. On the other hand, the $p_z$-orbitals overlap sideways with
regions of highest electron density above and below the graphene
plane and the energy dispersion of these $\pi$-bonds determines the
electronic transport properties of graphene materials.

\begin{figure}[!ht] 
\centering
\includegraphics[width=\textwidth]{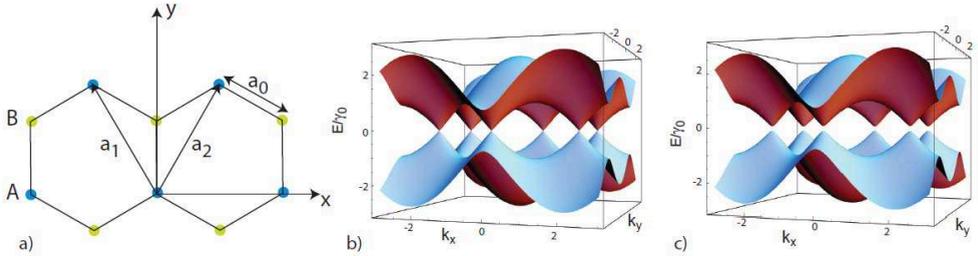} 
\caption{Panel (a) shows the crystal structure of monolayer graphene whose unit cell
contains two equivalent carbon atoms -$A$ and $B$. The 3D plot in (b) shows the energy dispersion of graphene (see Eq. \ref{Egraphene}). The valence and conduction band touch in 6 points, known as valleys. Whenever the onsite energy symmetry between the $A$ and $B$ sublattices is broken a band-gap opens in the energy dispersion of graphene as shown in Panel (d) -energy dispersion obtained considering $H_{AA} \neq H_{BB}$ with $H_{AA}=0.1 \gamma_0$ and $H_{BB}=-0.1 \gamma_0$.} \label{Russo_figure1}
\end{figure}

The hexagonal lattice of graphene is a composite lattice with two
carbon atoms in the unit cell -indicated by A and B, see Fig. \ref{Russo_figure1}a- and basis vectors:

\begin{equation}
\begin{cases}
\textbf{a}_{1} = - \sqrt{3}/2 a_{0} \hat{i} + 3/2 a_{0} \hat{j} \\
\textbf{a}_{2} = \sqrt{3}/2 a_{0} \hat{i} + 3/2 a_{0} \hat{j}.\\
\end{cases}
\label{1Lbasis}
\end{equation}

We consider the linear combination of atomic orbitals (LCAO) of the Bloch wavefunctions corresponding to the sublattice carbon atoms $A$ and $B$ of the form: 

\begin{equation}
\begin{array}{cc}
\phi_{i} = \frac{1}{\sqrt{N}} \sum_{n} e^{i \textbf{k} \cdotp \textbf{d}_{in}} \varphi_{i} (\textbf{r} - \textbf{d}_{in}), & i=A, B. \\
\end{array}
\label{Bloch}
\end{equation}

Where N is the number of cells considered, $\textbf{k}$ is the wave vector, $\textbf{d}_{in}$ is a lattice translation vector and $\varphi_{i}(\textbf{r}-\textbf{d}_{in})$ is the $p_{z}$ local atomic orbital. The $\pi$-orbitals electronic band structure is therefore a solution of the Schr\"odinger equation:

\begin{equation}
H \phi(\textbf{k})=E(\textbf{k})\phi(\textbf{k})
\label{Schroedinger}
\end{equation}

where $H$ is the Hamiltonian for an electron in the atomic potential given
by the atoms in the graphene lattice and $\phi(\textbf{k})$ is a linear combination of Bloch wavefunctions which for simplicity we can write in the form $\phi = a \phi_{A} + b \phi_{B}$, where $\phi_{A}$ and $\phi_{B}$ are given by Eq. \ref{Bloch} and $a$ and $b$ are two coefficients. To solve Eq. \ref{Schroedinger} we need to find the matrix elements $H_{ij}=<\phi_{i}|H|\phi_{j}>$ of the Hamiltonian and the overlaps between the Bloch wavefunctions $<\phi_{i}|\phi_{j}>$. We start from noticing that if the two carbon atoms forming the graphene sublattice are energetically equivalent, the onsite energies of the sublattice A and B are equivalent ($H_{AA}=H_{BB}$) and without loss of generality we can set this energy equal to zero. The solution of Eq.\ref{Schroedinger} is further simplified if we consider that the most significant hopping parameter is the first neighbour ($\gamma_{0}\approx 2.8 eV$, (\cite{CastroNeto09})) and that $H_{BA}$ is simply the complex conjugate of $H_{AB}$ ($H_{BA}=H_{AB}*$). Therefore we can calculate the integral $H_{AB}=<\phi_{A}|H|\phi_{B}>$ where each atom is surrounded by three neighbours with relative coordinates (1/3, 1/3), (1/3,-2/3), (-2/3, 1/3). The term $H_{AB}$ reads:

\begin{equation}
H_{AB} = \gamma_{0} (e^{i \textbf{k} \cdotp (\frac{\textbf{a}_{1}}{3} + \frac{\textbf{a}_{2}}{3})} + e^{i \textbf{k} \cdotp (\frac{\textbf{a}_{1}}{3} - \frac{2 \textbf{a}_{2}}{3})} + e^{i \textbf{k} \cdotp (-2 \frac{\textbf{a}_{1}}{3} + \frac{\textbf{a}_{2}}{3})}).
\end{equation}

We can now project the solution of Eq. \ref{Schroedinger} onto $<\phi_{A}|$ and $<\phi_{B}|$ to obtain the system:

\begin{equation}
\left\{
\begin{array}{c}
   a H_{AA} + b H_{AB} = E(\textbf{k}) a \\
   a H_{BA} + b H_{BB} = E(\textbf{k}) b \\
\end{array} \right.
\Rightarrow
\left\{
\begin{array}{c}
   b H_{AB} = E(\textbf{k}) a \\
   a H_{AB}* = E(\textbf{k}) b \\
\end{array} \right.
\label{1Lsystem}
\end{equation}

which has non-zero solutions for the coefficients $a$ and $b$ only if its secular determinant is zero. This condition leads to the energy dispersion of the graphene $\pi$-orbitals:

\begin{equation}
E(k_{x},k_{y})= \pm \gamma_{0} \sqrt{1+4 \cos(\frac{\sqrt{3}r}{2} k_{y}) \cos(\frac{r}{2} k_{x}) + 4 \cos^2 (\frac{r}{2} k_{x}) }
\label{Egraphene}
\end{equation}

where $r=a_{0} \sqrt{3}$. The energy distribution vanishes at six points in the reciprocal lattice space with coordinates $\pm 2 \pi / r(1/ \sqrt{3}, 1/3)$, $\pm 2 \pi / r(0, 2/3)$, $\pm 2 \pi / r(- 1/ \sqrt{3}, 1/3)$, see Fig. \ref{Russo_figure1}b. In these six K-space points the valence and conduction band touch one another, but only two of these points are independent. These are commonly indicated by K and K' and also known as valleys. The electronic states close to the Fermi level (E=0) are readily described by a Taylor expansion of the energy dispersion in Eq. \ref{Egraphene} at a chosen K point. This reveals that charge carriers in graphene are mass-less Dirac electrons obeying a linear energy dispersion: 

\begin{equation}
E(k)= \pm v_{F} |k|
\label{LinearDispersion}
\end{equation}

with $v_{F}=\sqrt{3} \gamma_{0} r / 2 \hbar$ the energy independent Fermi velocity (\cite{Novoselov005,Zhang05,Jiang07,ZhangY08}).

The absence of a band-gap in the energy dispersion of graphene implies that the conduction in this material cannot be simply switched on or off by means of a gate voltage which acts on the position of the Fermi level, limiting the use of graphene in conventional transistor applications. Indeed, even when the Fermi level in graphene devices is at $E=0$, the current in graphene is far from beeing completely pinched-off. However, the gapless energy dispersion of graphene is a consequence of the assumption that the electron onsite energy between the $A$ and $B$ sublattice carbon atoms are equal. Whenever $H_{AA}\neq H_{BB}$ a band-gap opens in the energy spectrum of graphene, see Fig. \ref{Russo_figure1}c. A viable way to experimentally engineer such a band-gap consists in growing and/or depositing graphene on a commensurate honeycomb lattice formed by chemically inequivalent atoms which ultimately will originate a difference in the onsite energy between the $A$ and $B$ sublattices. So far, hexagonal BN is considered to be one of the most promissing candidates for graphene band-gap engineering since it has an almost commensurate crystal structure to the one of graphene and it has two different elements in each sublattice (\cite{Giovannetti07}). However, despite the growing interest in graphene on h-BN, no experimental evidence has been reported yet of a band-gap opening in graphene on h-BN (\cite{Xue11}).

\begin{figure}[!ht] 
\centering
\includegraphics[width=10cm]{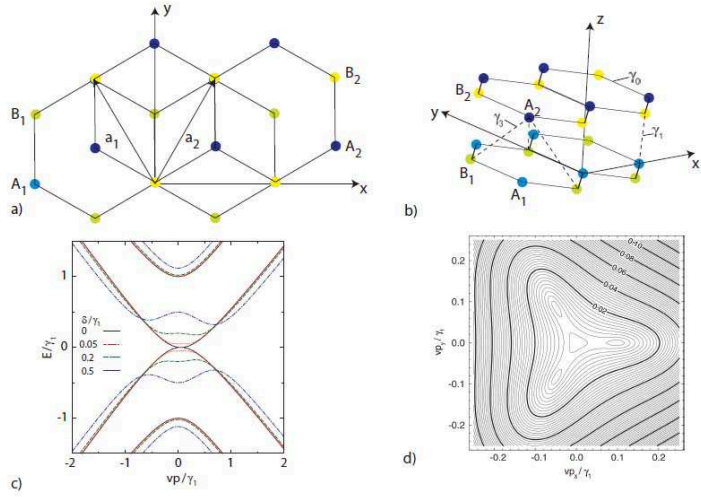} 
\caption{Panel (a) and (b) show respectively a top view and a 3D view of the crystal structure of AB-stacked bilayer graphene whose unit cell
contains four equivalent carbon atoms -$A_{1}, B_{1}, A_{2}$ and $B_{2}$. The hopping parameters $\gamma_0, \gamma_1$ and $\gamma_3$ are highlighted in the scheme. The graph in (c) shows the energy dispersion of bilayer graphene without an energetical asymmetry between the sublattices (continuous line) and with various values of the onsite energy difference (dashed lines)  (\cite{KoshinoNJP09}). When all the hopping parameters are considered in the tight binding calculation, the low energy dispersion of bilayers is not any more simply parabolic but 4 Dirac cones develop. Panel (d) shows an equi-energy contour plots of the lowest electron band of bilayer
graphene for equi-energy onsite simmetry case with the 4 Dirac cones clearly visible (\cite{KoshinoNJP09}).} \label{Russo_figure2}
\end{figure}

On the other hand, bilayer graphene offers a unique alternative to the problem of band-gap engineering for reasons which will become clear when considering the bilayer non-interacting tight binding description (\cite{McCann06}). The most common stacking of graphite planes found in nature is of Bernal type, where the $A$ atoms in one layer are alligned on top of $B$ atoms of an
adjacent layer. The unit cell of bilayer graphene consists of a
basis of four atoms labelled $A1$, $B1$, $A2$ and $B2$ belonging to
different atomic planes as indicated by the numerical index, see Fig. \ref{Russo_figure2}a and b.

Similarly to the case of a single layer we adopt the LCAO method with Bloch wavefunctions corresponding to the sublattice carbon atoms of the form: $\phi_{i} = \frac{1}{\sqrt{N}} \sum_{n} e^{i \textbf{k} \cdotp \textbf{d}_{in}} \varphi_{i} (\textbf{r} - \textbf{d}_{in})$ with $i=A1, B1, A2, B2$ and  $\textbf{d}_{in}$ the translation vector of the $i$ sublattice. For simplicity, we start by considering non-zero only the nearest neighbour coupling -i.e. $\gamma_{0}$ hopping from $A1$ to $B1$ and from $A2$ to $B2$ atomic sites- and $\gamma_{1} \approx 0.39 eV$ the interlayer coupling between $A2$ and $B1$
atoms (\cite{CastroNeto09}). We also assume that all the carbon atoms lattice sites are energetically equivalent -i.e. $H_{AiAi}=H_{BjBj}$ with $i,j=1,2$. In this case we consider the linear combination of Bloch wavefunctions of the form $\phi=a \phi_{A1} + b \phi_{B1} +c \phi_{A2} + d \phi_{B2}$, with $a,b,c$ and $d$ coefficients. The solution of the  Schr\"odinger equation is readily obtained projecting the solutions of Eq. \ref{Schroedinger} onto the states $<\phi_{i}|$ with $i=A1, B1, A2$, and $B2$. The equivalent system to Eq. \ref{1Lsystem} for the case of bilayer contains 4 equations, and imposing the condition that the secular determinant is zero leads to the set of the bilayer four bands:

\begin{equation}
E(k)=\pm \frac{\gamma_{1}}{2} \pm \sqrt{\frac{\gamma^2_{1}}{4}+(\hbar v_{F} k)^2}
\label{Ebilayer}
\end{equation}

with $v_{F}=\sqrt{3} \gamma_{0} r/2 \hbar$. The energy dispersion in Eq. \ref{Ebilayer} showes that the lowest energy conduction and valence band of bilayer graphene touch each other in each of the two K and K' valleys. The higher energy bands are instead shifted $\pm 0.4$ eV away from the $E=0$ Fermi level position. In the low energy limit, the energy dispersion in bilayers is parabolic (see Fig. \ref{Russo_figure2}c): 

\begin{equation}
E(k)= \pm \hbar^2 k^2 / 2 m*,
\label{2Lparabolas}
\end{equation} 

where m* is the charge particles effective mass $m*= \gamma_{1}/2v^2_{F}$. 

If we lift the energy degeneracy between the carbon sublattices on the two layers the gap-less energy dispersion of bilayers develops an energy gap around $E=0$ (\cite{McCann06,McCann2006a,Castro07,Ohta06,Oostinga08,Zhou08,ZhangLM08,Zhang09,Mak09,Kuzmenko09,Russo09,Xia10,Castro10,Taychatanapat10,Zou10}). The origin of this band-gap is readily understood when considering an onsite energy asymmetry $\Delta$ between the sublattices $A1$ and $B2$. In this case the energy dispersion opens an energy gap equal to $2 \Delta$ at each valley in the energy dispersion. This condition is easy to realize experimentally in double gated devices (see Section \ref{Experiments on double gated devices}) and it led to the discovery of a tuneable band-gap in bilayer graphene (\cite{Craciun10}).

\begin{figure}[!ht] 
\centering
\includegraphics[width=11cm]{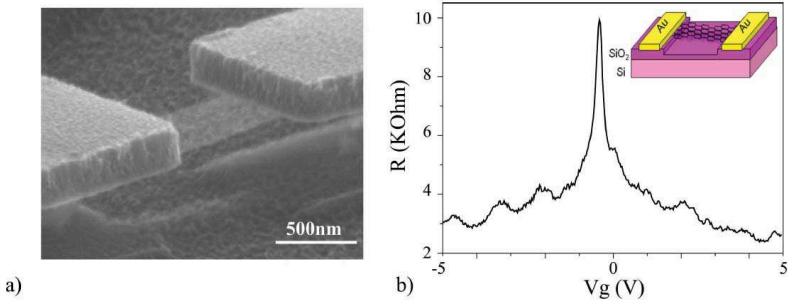} 
\caption{Panel (a) shows a scanning electron microscope image at an angle of 60\textdegree of a suspended 2-terminal bilayer graphene device, obtained by wet-etching of the $SiO_2$ substrate. The graph in (b) is a low temperature (T=0.3K) electrical transport measurement of the suspended bilayer (charge mobility of $\approx 20000 cm^2/Vs$). A series resistance of 1.2KOhm has been subtracted to account for contact resistance (\cite{Russo10}) and for the resistance of the electric lines of the cryostat.} \label{Russo_figure3}
\end{figure}

Recent advances in ultraclean suspended graphene devices made it possible to study details of the energy dispersion of bilayer in the vicinity of E=0 and at energy scales much smaller than the nearest neighbour hopping, see Fig. \ref{Russo_figure3}. In this case, the tight-binding description which includes higher order hopping parameters reveals that the energy dispersion of bilayers has a trigonally deformed dispersion (see Fig. \ref{Russo_figure2}d) with 4 touching points in each valley (\cite{McCann2006a,Mikitik08,KoshinoNJP09}). In the vicinity of each touching
point the energy dispersion is linear (in the low energy 4 Dirac cones appear) and not parabolic as
predicted within the nearest neighbour hopping approximation.
Therefore, depending on the energy scale, the Dirac fermions of
bilayer graphene can lose or gain an effective mass.

The topological discontinuity of the Fermi surface which occurs when crossing
from the Dirac cones to parabolic energy dispersion is a
Lifshitz transition (\cite{Lifshitz01}). It is rather obvious that a
discontinuous change in the topology of the Fermi surface will be
reflected in singularities of the thermodynamic and kinetic
observables of the system. Since in bilayer graphene transistor
devices the Fermi level can be continuously tuned by a gate
voltage, this system is ideal to study the occurrence of the
Lishitz transition and its implication on the transport properties (\cite{Lemonik10}).
Currently, fundamental questions such as the stability of these 4 Dirac cones against electron-electron
interactions and/or mechanical deformations are at the focus of both
theoretical and experimental research.

\begin{figure}[!ht] 
\centering
\includegraphics[width=0.8\textwidth]{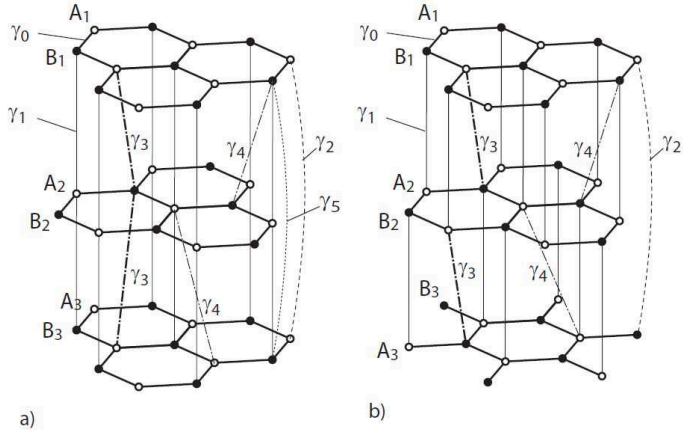} 
\caption{Panels (a) and (b) show respectively the crystal structure of Bernal stacked  and rhombohedral stacked trilayer graphene with the complete set of tight binding parameters. } \label{Russo_figure4}
\end{figure}

For multilayers of more than two layers,
there are two known structures
called ABA (hexagonal or Bernal) and ABC (rhombohedral)
with different stacking manners as shown in Fig. \ref{Russo_figure4}a and b.
For ABA multilayer graphene, the Hamiltonian 
can be approximately decomposed 
into a superposition of the monolayer-like and bilayer-like
subsystems (\cite{Koshino07,Koshino08}). Specifically, $2M$-layer graphene
contains $M$ bilayer-like subbands, while
$(2M+1)$-layer graphene has $M$ bilayer-like subbands
and one monolayer-like subbands.
The subbands appearing in $N$-layer graphene 
are characterized by a single parameter (\cite{Guinea06,Koshino07,Koshino08}):

\begin{eqnarray}
\lambda = 2 \cos \frac{n \pi}{N+1},
\end{eqnarray}

where $n = 1,2,\cdots,[(N+1)/2]$ is the subband index 
and $[(N+1)/2]$ is the maximum integer which does not exceed $(N+1)/2$.
The sub-Hamiltonian of $\lambda \neq 0$
is equiavelent to that of bilayer graphene
with the nearest interlayer coupling parameters
multiplied by $\lambda$,
while that of $\lambda =0$, appearing in odd $N$ only, 
becomes that of monolayer graphene.
For instance, the trilayer-graphene comprises 
six bands of which two have linear (monolayer-like)
dispersion and four have parabolic disperion (bilayer-like).
On the other hand, ABC multilayers have quite different electronic structures: the low-energy spectrum contains only a pair of conduction and 
valence bands touching at zero energy (\cite{Guinea06,Min08,Koshino09,Koshino10}).  
These two bands are 
given by the surface states localized at outer-most 
layers, and the dispersion is roughly proportional to $k^N$,
and becomes flatter as $N$ increases.

\begin{figure}[!ht] 
\centering
\includegraphics[width=0.8\textwidth]{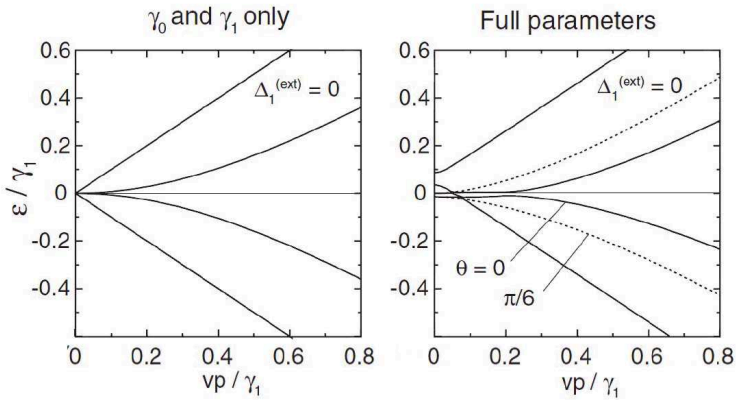} 
\caption{The graphs show the self-consistently calculated band structures in trilayer
graphene near the K point respectively in the first neighbour approximation (i.e., $\gamma_0$ and $\gamma_1$ only) and in the full parameter model, adapted from \cite{Koshino09}.} \label{Russo_figure5}
\end{figure}

The response of ABA multilayer graphene
to the gate electric field perpendicular to the layers
is unique and qualitatively different from 
that of AB-bilayer graphene (\cite{Craciun09,Russo09}). In trilayer graphene, for example,
the gate field breaks mirror reflection
symmetry with respect to the central layer and hybridizes the linear and
parabolic low-energy bands, leaving the semi-metallic band overlap
near zero energy (\cite{Craciun09,Koshino09}). Only the tight
binding model based on the full set of hopping integrals (i.e.
$\gamma_{0}$ to $\gamma_{5}$ see Fig. \ref{Russo_figure5}) can succesfully describe
the experimentally observed band-overlap (\cite{Russo09,Craciun09}. The tuneable
semimetallic nature of trilayers together with the tuneable
band-gap in bilayers demonstrates that graphene materials are
characterized by a unique range of physical properties not found
in any other known material system. Theoretically, a similar band overlapping is expected in thicker ABA multilayers as well, though very little is known experimentally.

Finally, graphene multilayers can also be constructed by carefully transferring layer by layer individual graphene sheets on a substrate. The rise of this new family of materials, the meta-few layers ($\mu \epsilon \tau \alpha$ = "beyond" the few layer materials), offers the unique possibility to control the hopping parameters by stacking engineering. In particular, the interlayer hopping parameters are responsible for the rich low-energy band dispersion of these materials. Therefore, the ability to control these hopping parameters by means of graphene stacking engineering -for instance making thin tunnel barriers between two subsequent graphene layers- holds the promise for unprecedented functionalities as compared to natural few-layers. This layer-by-layer engineering of graphene materials has already been successfuly employed to create transparent electrodes in organic solar cells, and an efficiency similar to the ITO electrodes was recently demonstrated (\cite{Wang11}).

The physics of graphene has expanded at a rapid pace mainly thanks to the easily accessible electronic properties in simple transistor geometries, see Fig. \ref{Russo_figure6}a. In these devices, metallic contacts inject charge carriers into the conductive graphene channel, whereas the Fermi level of graphene is continuously driven from the valence to the conduction band by means of a gate voltage. As the Fermi level is driven inside the conduction (valence) band, the conductivity increases with increasing the concentration of electrons (holes) induced by the gate voltage. At the touching point between the valence and conduction bands the Fermi level crosses the zero density of states point -i.e. the Dirac point- where the conductance reaches its minimum value. Indeed, despite the density of sates vanishes at the Dirac point the conductivity remains finite with a minimum value of $ \approx 4e^2/ \pi h$ for the ballistic transport regime (\cite{Novoselov005,Katsnelson06,Tworzydlo06,Geim07}). Theoretically a finite conductivity is expected for ballistic electrical transport in graphene at the neutrality point assisted by evanescent wave propagation. This evanescent wave propagation gives rise to a unique non-monotonous dependence of the Fano factor on the charge density in shot-noise (\cite{Tworzydlo06,Danneau08a,Danneau08b,Danneau09}). The predicted minimum conductivity $ 4e^2/ \pi h$ has only been experimentally observed in devices with a short graphene channel. So far, most of the experimental studies on the minimum of conductivity have been conducted in supported graphene -e.g. graphene on SiO2 substrate. Ultrapure suspended graphene devices do not suffer of the presence of substrate-related disorder and are therefore the ideal candidate to address the physics governing electrical transport at the minimum of conductivity.

Indeed, the presence of disorder such as adatoms or molecules, ripples of the graphene sheet, atomic vacancies and topological defects are expected to affect the electronic properties of graphene such as the conductivity and charge carrier mobility. In particular, close to the Dirac point, charged impurities create electron/hole puddles which dominate the charge transport properties of graphene (\cite{Martin08}). For small energies around the Dirac point recent theoretical advances also pointed out that strong short-range interaction caused by scattering off adatoms and molecules on the surface of graphene can actually be the dominant source of disorder limiting the charge carrier mobility. During the fabrication process of transistor devices, graphene is exposed to environmental conditions, it is therefore likely that for instance hydrocarbons are adsorbed on the surface of graphene. Whenever an hydrocarbon covalently bonds to the graphene, the $2p_{z}$ delocalized electrons are localized into a $\sigma$-bond, i.e. a covalently bond hydrocarbon effectively act as a vacancy. The charge carrier scattering off the resonant states induced in the vicinity of the Dirac point together with scattering off charged impurities are likely to play a dominant role in graphene devices, and this is currently at the focus of both theoretical and experimental research (\cite{Peres10,Ferreira11}). 

\section{Experimental observation of gate tuneable band structure in few-layer graphene}\label{Experiments on double gated devices}

One of the most remarkable physical property of graphene materials is the ability to reversibly tune the band structure of these systems simply by means of an external electric field (\cite{Craciun10}). In standard semiconducting materials a precise value of the band-gap is engineered during the growth process, therefore the value of this energy gap cannot be reversibly controlled in situ in a device. Few-layer graphene is the only known class of materials to exhibit a gate tuneable band structure and this unprecedented property paves the way for devices with novel functionalities (\cite{Craciun10}). 

Different experimental approaches have been implemented to address the band structure of graphene materials. Respectively, pioneering charge transport experiments (\cite{Castro07,Oostinga08,Craciun09}) followed by photoemission spectroscopy (\cite{Ohta06}) and infrared spectroscopy (\cite{ZhangLM08,Zhou08,Zhang09,Mak09}) have highlighted complementary aspects of the energy dispersion of few-layer graphene with different number of layers. 

\begin{figure}[!htb]	
\centering
\includegraphics[width=\textwidth]{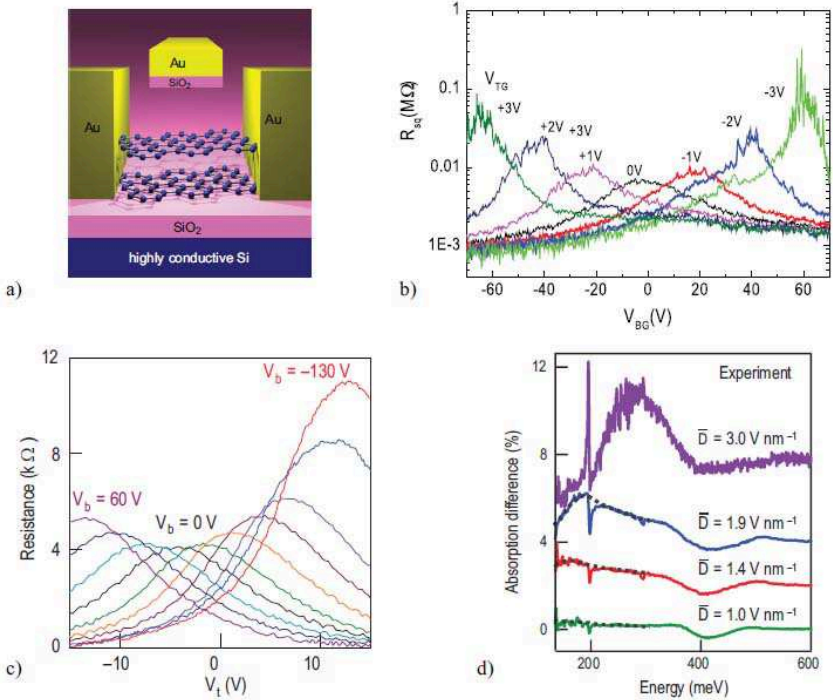} 
\caption{(a) Schematic picture of a double gated bilayer graphene device. (b) Square resistance Rsq of bilayer graphene as a function of back gate voltage ($V_{BG}$) measured for different fixed values of the top gate voltage ($V_{TG}$) at T=300mK. The position of the Fermi level and the applied perpendicular electric field are controlled by $V_{BG}$ and $V_{TG}$, adapted from \cite{Russo09}. (c) the square resistance of a bilayer device versus $V_{t}$ for different values of the back gate voltage $V_b$ and (d) shows the corresponding infrared abosorption spectra, adapted from \cite{Zhang09}.$\overline{D}$ represents the average value between the top- and back-gate electric displacements applied to the bilayer.}\label{Russo_figure6}
\end{figure}

Possibly the best device geometry in which to address the electric field tuneability of the FLGs energy dispersion is a double gated design, where the graphene materials are sandwitched between a top- and a back-gate, see Fig. \ref{Russo_figure6}a. Double gated geometries have a dual valency, that is they offer a simple way to independently and continuously control in-situ both the band structure and the Fermi level by means of gate voltages (\cite{Craciun10}). A typical device layout comprises a source and a drain contact to a graphene flake exfoliated onto $SiO_2 /Si$ substrate -which serves as a back gate- and a nanofabricated top gate. Since the conductance is determined by the features of the energy bands in the thermal shell $k_B T$ around the Fermi energy ($\varepsilon_{F}$), any modification of the conductivity in response to the perpendicular electric field is purely a consequence of the changes in the energy dispersion. This intimate relation between conductivity and energy dispersion has allowed the discovery that bilayer graphene is the only known semiconductor with a gate tuneable band-gap (\cite{Zhang09}) and conversely that trilayer graphene is the only known semimetal with a gate tuneable conduction and valence bands overlap (\cite{Craciun09}).

A perpendicular electric field applied onto the few-layer graphene materials breaks the energetic symmetry between the planes of these multilayer systems (see Section \ref{Electronic properties of FLGs}). This asymmetry is then reflected in the energetic inequivalence between carbon atoms belonging to different sublattices -which in multilayer graphene belong to different layers. Experimentally it was observed that the in-plane electrical transport properties of each specific FLG thickness change in a unique way in response to a finite external electric field ($E_{ex}$). In all cases the resistance exhibits a maximum whose value and position in gate voltage depend on the voltage applied to the gate on which a fixed potential is applied during the measurement. 

In bilayers whenever $E_{ex} \neq 0$ V m$^{-1}$ the maximum of square resistance  ($R_{sq}^{max}$) displays a metal-to-insulator transition consequence of the opening of an electric field induced gap in the energy dispersion. Once the energetic equivalence between the sublattices is restored (i.e. $E_{ex}=0$), the energy gap reduces to zero. Consistently, charge transport experiments have reported large on/off ratios of the current in double gated graphene bilayers when the Fermi level crosses from the conduction (valence) band through the band-gap, see Fig. \ref{Russo_figure6}b and c. However, the values estimated for the band-gap from transport experiments are systematically much smaller than the theoretically predicted energy gap. Typically in transport a \textsl{mobility gap} is observed on an energy scale of a few meV for an average perpendicular electric displacement field of $\overline{D} = 2 V/nm$, whereas a band gap of 200 meV is theoretically expected. Furthermore, the specific temperature dependence of $R_{sq}^{max}$ measured in bilayer graphene is incompatible with a simple thermally activated transport over a band gap, but it exhibits the functional dependence typical of variable range hopping due to a finite sub-gap density of states (\cite{Oostinga08,Russo09,Taychatanapat10,Zou10,Yan10}). 

The dichotomy of a \textsl{mobility gap} in transport experiments and the theoretically expected energy gap is currently fuelling both theoretical and experimental discussions. Though several possible explanations have been put forward, transport studies in different geometries demonstrate that the temperature dependence of the conductance in bilayer graphene in the diffusive regime can be explained by the parallel of a thermally activated current over the energy-gap and a variable range hopping current through the disorder induced sub-gap states (\cite{Zou10,Yan10,Taychatanapat10}). 

The first direct observation of a gate-tuneable energy gap in bilayers was reported in infrared spectroscopy experiments (\cite{ZhangLM08,Zhou08,Zhang09,Mak09}). This technique is mostly sensitive to band to band transitions, therefore  it is not affected significantly by transitions in the sub-gap energy range. Whenever a band-gap is open in bilayer graphene, the infrared absorption displays a highly intense peak in the absorption spectra corresponding to the transition of charge particles from the top of the valence band to the bottom of the conduction band, see Fig. \ref{Russo_figure6}d. This peak in the infrared absorption spectra corresponds to the energy gap between the valence and conduction band and it has a pronounced gate tuneability. In this way an electric field induced gap ranging from 0 meV up to 250 meV at $\overline{D} = 3V/nm$ has been reported, which is consistent with theoretical predictions. At the same time, transport experiments in the very same devices show a small increase of the maximum resistance as a function of $E_{ex}$ confirming the presence of a large disorder induced sub-gap density of states.

\begin{figure}[!ht] 
\centering
\includegraphics[width=\textwidth]{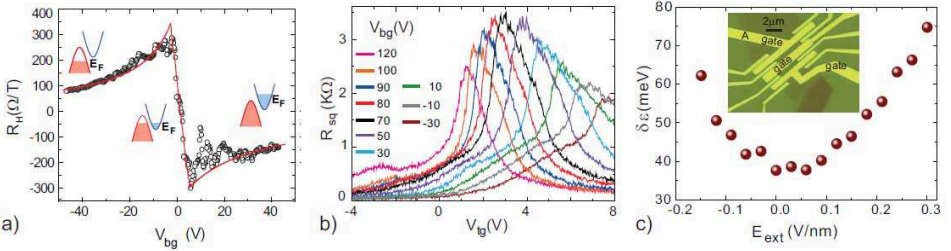} 
\caption{ (a) shows the Hall coefficient as a function of back gate voltage $V_{BG}$ (open circles) for a fixed perpendicular magnetic field of 9T at 50mK for a trilayer graphene device. The red curve is a fit. The insets depict schematically the position of the Fermi level ($\varepsilon_{F}$) at different values of $V_{BG}$. The graph in (b) shows a 4-terminal resistance measurement of the trilayer device in the inset of (c) \textsl{versus} top gate and for different values of back gate. (c) shows the electric field dependence of the band overlap $\delta \varepsilon$ for the same trilayer device. All panels are adapted from \cite{Craciun09}} \label{Russo_figure7}
\end{figure}

Contrary to bilayers, ABA-stacked trilayers display a decrease of $R_{sq}^{max}$ with increasing $E_{ex}$, see Fig. \ref{Russo_figure7}b (\cite{Craciun09,Russo09}). The overall electric field dependence of the resistance of trilayers can be explained adopting a two band model with an electric field tuneable band overlap between the conduction and valence band ($\delta \varepsilon$). To this end trilayer graphene is the only known semimetal with a gate tuneable band overlap, see \ref{Russo_figure7}c . This unique property was independently demonstrated by magneto-transport experiments of the Hall coefficient ($R_{H}$), see Fig. \ref{Russo_figure7}a. In particular, $R_{H}$ measured at a fixed external perpendicular magnetic field displays a characteristic sign reversal corresponding to the cross-over between different types of charge carriers involved in the conduction (electrons and holes). The back gate voltage range over which $R_{H}$ changes sign gives a band overlap $\delta \varepsilon \approx 28$ meV, see Fig. \ref{Russo_figure7}a.

To date little is known experimentally on thicker few layer graphene materials with more than 3 layers. Recent infrared spectroscopy experiments address the evolution of the electronic properties from the one of mass-less Dirac electrons in a single layer to the massive particles of bulk graphite, presenting a systematic study from 1 up to 8 Bernal stacked graphene layers. Measurements of infrared conductivity show that the key features of
the 2D band structure of few-layer graphene can be achieved on the basis of zone folding of the 3D graphite bands (\cite{Mak10}). However, so far the electron transport properties of these thicker few-layer graphene materials is largely unexplored, preventing us from identifying the best suited thickness of few-layer graphene for a given application.

\section{Landau level structures in few-layer graphene}\label{Quantum Hall}

When the charge carriers travelling in few-layer graphene experience a perpendicular magnetic field (B), their trajectories are bent due to the Lorentz force. In the quantum regime, these cyclotron orbits give rise to discrete energy levels known as Landau levels. The precise sequence of these Landau levels reflects the nature of the charge carriers in the few-layer graphene under consideration. In particular, the Landau levels sequence for single layer graphene is intimately related to the mass-less nature of the graphene Dirac fermions (see Eq. \ref{LinearDispersion}) and it is very different from what is known in conventional two-dimensional electron gases. 
 
The Landau level energies of monolayer graphenes are given by 
(\cite{McClure1956a}):

\begin{eqnarray}
E_{n} = \hbar \omega_B {\rm sgn}(n) \sqrt{|n|} && (n=0, \pm 1, \pm 2, ..),
\label{LLgraphene}
\end{eqnarray}

with $\hbar \omega_B= \sqrt{2 \pi v_{F}^2 e B}$. Each level is four fold
degenerate, that is spin and valley degenerate. Due to the linear energy
dispersion (see Eq. \ref{LinearDispersion}), the energy spacing between the Landau
levels is proportional to  $\sqrt{B}$ rather than B as in usual
two dimensional systems. At a fixed value of external magnetic field the
energy gap between Landau levels in graphene are much larger than the
corresponding gaps opened in other 2DEGs (for B=1T the energy gap
between n=0 and n=1 in graphene is approximately 35 meV). Another unique
feature of the graphene Landau level spectrum is the existence of a zero
energy level. This causes the half-integer quantization of Hall
conductivity per spin and valley, 
and is also responsible for the huge diamagnetic susceptibility 
characteristic of this system.
Specifically, the orbital susceptibility has a strong singularity at
band touching point (Dirac point), which 
at zero temperature is expressed 
as a function of Fermi energy $\varepsilon_F$ as:

\begin{equation}
\chi(\varepsilon_F) = -g_vg_s \frac{e^2 v^2}{6\pi c^2} \delta(\varepsilon_F),
\end{equation}
\label{eq_chi_mono}

where $g_v=g_s=2$ are respectively the spin and valley degeneracies (\cite{McClure1956a}).

\begin{figure}[!ht] 
\centering
\includegraphics[width=\textwidth]{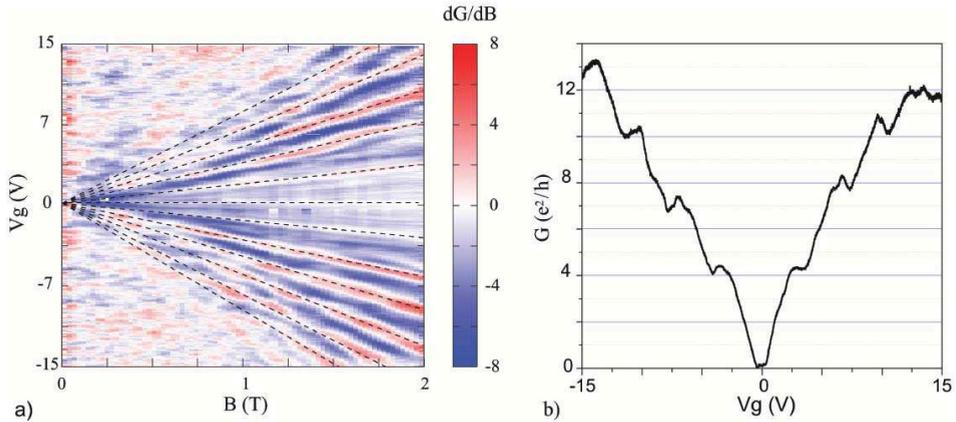} 
\caption{Panel (a) is a color coded plot of dG/dB \textsl{versus} gate voltage for the magnetic field range of 0T to 2T and at T=0.3K measured in the suspended bilayer device of Fig. \ref{Russo_figure3}. As the magnetic field is increased in addition to an insulating state at filling factor $\nu=0$, also the quantum Hall plateaus at $\nu= \pm4, \pm8, ...$ are visible (dashed lines are a guideline for the eyes). The graph in (b) is a plot of the conductance \textsl{versus} gate voltage for B=2T (T=0.3K) showing the bilayer Hall quantization sequence.} \label{Russo_figure8}
\end{figure}

The low-energy Landau levels dispersion of bilayer graphene is 
approximately given by the relation (\cite{McCann2006a}):

\begin{eqnarray}
E_{sn}= s \hbar \omega_c \sqrt{n(n-1),} && (s= \pm, \,\, n=0, 1, 2, ..)
\end{eqnarray}
\label{LLbilayer}

with $\omega_c=e B/ m^*$ the cyclotron frequency associated with the
effective mass $m^*$ of bilayer graphene. The Landau levels energy
spacing is now linear in B owing to the usual quadratic energy
dispersion of bilayers, see Fig. \ref{Russo_figure8}a. The two lowest levels of $n=0$ (per spin
and valley) appear at zero energy. This 
amounts to 8-fold degeneracy in total and
causes doubling of the Hall conductivity jump at zero electron
density, see Fig. \ref{Russo_figure8}b. 
The orbital susceptibility for small Fermi energy becomes (\cite{Safran1,Koshino07}):

\begin{equation}
\chi(\varepsilon_F) = - 
g_v g_s \frac{e^2v^2}{4\pi c^2 \gamma_1}
\left(
- \ln\frac{|\varepsilon_F|}{\gamma_1} 
\right),
\end{equation}
\label{eq_chi_bi}

which has a logarithmic singularlity 
in contrast to the delta-function in monolayer graphene.
The sigularity is weaker than in monolayer,
since the Landau level spacing is narrower 
so that the total energy gain in magnetic field at $\varepsilon_F =0$ becomes 
smaller.
When increasing the magnetic field amplitude, the energy of the particular
Landau level other than zero-energy levels crosses over from linear B to
$\sqrt{B}$ , in accordance with the crossover of the zero-field 
dispersion from linear to quadratic. 

Quite recently, the experimental observation
of the magnetotransport and the quantum Hall effect
is revealing yet a rich scenario of Landau level spectrum in trilayer graphene (\cite{Zhang2011,Bao2011,Taychatanapat2011,Kumar2011}).
For ABA multi-layer graphenes,
the Landau level spectra can be
again decomposed into a superposition of the monolayer and bilayer
subsystems as introduced in Sec. \ref{Electronic properties of FLGs}. 
In this case, the physical properties in magnetic fields, such as Hall conductivity and
the magnetic susceptibility can be expressed as the summation over
components of subsystems (\cite{Koshino07,Koshino08}). 
In trilayer graphene, for example,
the spectrum is composed of bilayer and monolayer Landau levels,
resulting in a 12-fold degeneracy at zero energy.
The effect of the next-nearest interlayer couplings,
such as $\gamma_2$ and $\gamma_5$ (see Fig. \ref{Russo_figure4} a),
are often neglected in the simplest approximation,
but become particularly important
for the low-energy spectrum near the charge neutrality point (\cite{Koshino2011}).
For trilayer, the 12-fold degeneracy is lifted by 
those couplings, causing a qualitative change in the 
quantum Hall plateau structure.
The Landau spectrum of ABC multilayers
is quite different from ABA's,
where the pair of low-energy
flat-bands gives the Landau level sequence (\cite{Guinea06,Koshino09}):

\begin{eqnarray}
 E_{sn} = s \frac{(\hbar\omega_B)^N}{\gamma_1^{N-1}}
\sqrt{n(n-1)\cdots (n-N+1)},
&& (s= \pm, \,\, n=0, 1, 2, ..),
\end{eqnarray}

where $N$ is the number of layers.
Including valley and spin degree of freedom,
$4N$-fold degenerate Landau levels appear
at zero energy.

\section{Conclusions}

The gate tuneable band structure of FLGs is an unprecedented physical property which paves the way to conceptually novel physical phenomena. For instance, an asymmetry induced by a perpendicular electric field applied onto bilayer graphene, not only opens a gap but it also affects the pseudospin of the charge carriers \cite{Min08,San-Jose09}. This pseudospin characterises the layer degree of freedom, and it constitutes an additional quantum number for the charge carriers. Recent theoretical schemes propose the use of the pseudospin for new devices in which an on/off state of the current is attained respectively for parallel and antiparallel pseudospin configurations in the bilayer. In these pseudospin-valve devices the polarity of the electric field acting on bilayer graphene plays a similar role as the magnetic field in spin-valve devices. This is the emerging field of pseudo-spintronics.

\section{Acknowledgements}
S.R. and M.F.C. acknowledge financial support by EPSRC (Grant no. EP/G036101/1 and no. EP/J000396/1). S.R. acknowledges financial support by the Royal Society Research Grant 2010/R2 (Grant no. SH-05052). M.Y. acknowledges financial support by Grant-in-Aid for Young Scientists A (no. 20684011) and ERATO-JST (080300000477). S.T. acknowledges financial support from Special Coordination Funds for Promoting Science and Technology (NanoQuine), JST Strategic International Cooperative Program and MEXT Grant-in-Aid for Scientific Research on Innovative Areas (21102003).

\end{document}